\documentclass[a4paper,fleqn,usenatbib]{mnras}

\usepackage{newtxtext,newtxmath}

\usepackage{float} 
\usepackage{graphicx}	
\usepackage{amsmath}	
\usepackage{amssymb}	

\title[RLOF in binaries and anomalous unevolved stars]{Mass transfer of low-mass binaries and chemical anomalies
among unevolved stars in globular clusters}

\author[Wei et al.]{
Dandan Wei,$^{1,2,3,4}$\thanks{E-mail: wdd@ynao.ac.cn}
Bo Wang,$^{1,2,4}$ 
Xuefei Chen,$^{1,2,4}$
Hailiang Chen,$^{1,2,4}$  
Lifang Li,$^{1,2,4}$ \and
and Dengkai Jiang$^{1,2,4}$\thanks{E-mail: dengkai@ynao.ac.cn}
\\
$^{1}$ Yunnan Observatories, Chinese Academy of Sciences, Kunming 650216, China;\\
$^{2}$Key Laboratory for the Structure and Evolution of Celestial Objects, Chinese Academy of Sciences, Kunming 650216, China\\
$^{3}$University of Chinese Academy of Sciences, Beijing 100049, China\\
$^{4}$Center for Astronomical Mega-Science, Chinese Academy of Sciences, Beijing 100012, China
}

\date{Accepted XXX. Received YYY; in original form ZZZ}

\pubyear{2019}

\begin{document}

\label{firstpage}
\pagerange{\pageref{firstpage}--\pageref{lastpage}}
\maketitle

\begin{abstract}
While it is well known that mass transfer in binaries can pollute the surfaces of the accretors, it is still unclear whether this mechanism can reproduce the observed chemical inhomogeneities in globular clusters. We study the surface abundances of the accretors in low-mass binaries, as a first step towards understanding whether mass transfer in low-mass binaries is one of the potential origins of the aforementioned abundance anomalies in globular clusters.  We use the Modules for Experiments in Stellar Astrophysics code to calculate binary evolutionary models with different initial donor masses between 0.9 and 1.9\,$ \rm {M}_\odot$ for an initial metallicity of Z = 0.0034. The results show that in some low-mass binary systems, the accretors exhibit peculiar chemical patterns when they are still unevolved stars, e.g. C and O depletion; Na and N enhancement; constant Mg, Al and C+N+O. The abundance patterns of the accretors are significantly different from their initial abundances (or that of normal single stars), and can match the observed populations. These abundance patterns strongly depend not only on the initial parameters of binaries (donor mass, mass ratio, and orbital period), but also on the assumptions regarding mass-transfer efficiency and angular momentum loss. These results support the hypothesis that mass transfer in low-mass binaries is at least partly responsible for the unevolved anomalous stars in globular clusters. More work on binary evolutionary models and binary population synthesis is required to fully evaluate the contribution of this scenario.
\end{abstract}




\section{Introduction}
\label{introduce}

Galactic globular clusters (hereafter GCs) have been thought as good examples of simple stellar populations. All stars in one globular cluster have the same age, initial chemical composition, and their masses follow an initial mass function (IMF) \citep{Gratton2012AA}. However, a large number of spectroscopic and photometric studies provided  evidence for multiple stellar populations in GCs, see e.g. the reviews by  \cite{1994PASPKraft}, \cite{2004ARA&AGratton}, \cite{2016RAALi} and \cite{Bastian2018ARA&A}. For example, the splits (or a broadening) in the main sequence (MS), red giant branch (RGB) and horizon branch have been found in the color magnitude diagram (CMD)  \citep{Carretta2007APJ, Piotto2007ApJ, Milone2009A&A}. In addition, the abundances of light elements show inhomogeneities among stars within the same globular cluster: the anti-correlations of C-N, Na-O, and Mg-Al \citep{Carretta2009aA&A, Carretta2009bA&A}, while the abundances of Fe-group elements and the sum of C, N and O are approximately constant in many GCs \citep{Dickens1991Natur, Ivans1999AJ, Villanova2010ApJ}. But the two phenomena mentioned above maybe not directly related \citep{Mucciarelli2014,Martocchia2017MNRAS,Martocchia2018MNRAS}. It seems that the complexities of CMD in GCs are caused dy stellar structure of stars, such as stellar rotation \citep{Bastian2009MNRAS,D'Antona2015MNRAS,Bastian2018ARA&A,Kamann2018MNRAS,Milone2018MNRAS,2019MNRASMartocchia}. In this paper, we focus on the chemical anomalies in multiple stellar populations. A few years ago, such chemical patterns had only been found in clusters older than 10\,Gyr, but chemical anomalies were also detected in younger ($\geq$ 2\,Gyr) clusters now \citep{2019MNRASMartocchia}. As one of the most important tracers of chemical inhomogeneities, the Na-O anti-correlation exists in nearly all observed GCs \citep{Carretta2009aA&A, Carretta2009bA&A}. 


These abundance patterns could be explained by invoking the material processed by hydrogen burning, including the CNO, NeNa and MgAl cycles \citep{Denisenkov1990SvAL,Langer1995PASP}. But it is still an open question how this material is brought to the surface of stars presently observed in GCs. The evolutionary mixing was firstly proposed to reproduce the observed C-N anti-correlation as the normal star evolves to RGB star \citep{Denisenkov1990SvAL}. But this scenario can not explain similar abundance anomalies found in unevolved or scarcely evolved stars  \citep{Gratton2001A&A,Ramirez2002AJ,Bedin2004ApJ,Carretta2004A&A,Gratton2004MmSAI,Piotto2005ApJ,
Piotto2007ApJ,D'Orazi2010ApJ}. 

The most popular scenario is self-enrichment scenario. This scenario holds the view that more than one star formation episode has occurred within each GCs, and the second-generation stars are made up of the ejecta from a fraction of the first-generation stars (usually called ``the polluters''). A number of polluter candidates have been proposed, including fast rotating massive stars \citep{Decressin2007A&A}, supermassive stars \citep{Denissenkov2014MNRAS}, massive interacting binaries \citep{Mink2009A&A}, intermediate-mass close binaries \citep{2012A&AVanbeveren} and asymptotic giant branch stars  \citep{Decressin2009A&A,2010MNRASD'Ercole}.

However, these self-enrichment scenarios above have encountered a difficulty. The fraction of anomalous stars (the second-generation stars) observed within GCs is larger than that of the primordial population (the first-generation stars) \citep{ Bastian2018ARA&A, 2019A&ALarsen}. The currently proposed polluters in these self-enrichment scenarios are mainly intermediate-mass or massive stars with mass more than 2.5\,$\rm {M}_\odot$. So according to the standard stellar IMF, there is not enough material processed and ejected by these polluters to explain the fraction of the anomalous stars, which is the so called ``mass budget problem'' \citep{Prantzos2006A&A, Renzini2015MNRAS}. One solution to solve this problem is that GCs were assumed 10--100 times more massive initially, and a large fraction of first-generation stars (normal stars) were lost during their evolution \citep{Bastian2018ARA&A, 2008MNRASD'Ercole, 2010MNRASD'Ercole, 2010ApJVesperini}. Another possible explanation is that additional scenarios could operate simultaneously within the same cluster, including peculiar evolution of individual stars \citep{Gratton2012AA}, early disc accretion scenario \citep{Bastian2013MNRAS}, reverse population order for GCs formation scenario \citep{2009MNRASMarcolini}, and so on. 

It should be noted that the mass transfer in binaries (e.g. stable Roche lobe overflow) is another possible mechanism to reproduce star-to-star abundance variations in unevolved stars \citep{Gratton2012AA, 2015AJMilliman}. GCs are known to contain a certain fraction of binaries \citep{Milone2012A&A}, and the presence of the companion can alter the evolution of an already formed star by mass transfer. So one of the consequences of mass transfer in binaries is that surface abundances of the accretors may be different from that of single stars. This mechanism has been used to reproduce many abundance anomalous stars (including unevolved stars) observed in the field, such as fluorine-enrichment stars \citep{Lugaro2008A&A}, barium stars \citep{1988A&ABoffin, Han1995MNRAS}, carbon-enhanced metal-poor stars \citep{Abate2013A&A}. Besides, mass transfer in binaries has been one of the important explanations of many peculiar objects in globular clusters, e.g. blue stragglers and contact binaries \citep{Rucinski2000AJ,Knigge2009Natur,Jiang2017ApJ}. But it remains uncertain whether the mass transfer in binaries can reproduce the abundance anomalies observed in GCs (e.g. Na-O anti-correlation).

In this paper, we study the surface abundances of the accretors in low-mass binaries as a first step towards understanding whether this scenario is one of the potential origins of the abundance anomalies in GCs. In Section \ref{methods}, we show the formation of anomalous stars through the stable  Roche lobe overflow (RLOF) scenario, and introduce the methods of computation. The results are presented in Section \ref{results}. Finally, we give the discussion and conclusions in Section \ref{discussion}.

\section{Methods}
\label{methods}
\begin{figure}

\begin{center}
 \includegraphics [width=0.25\textwidth]{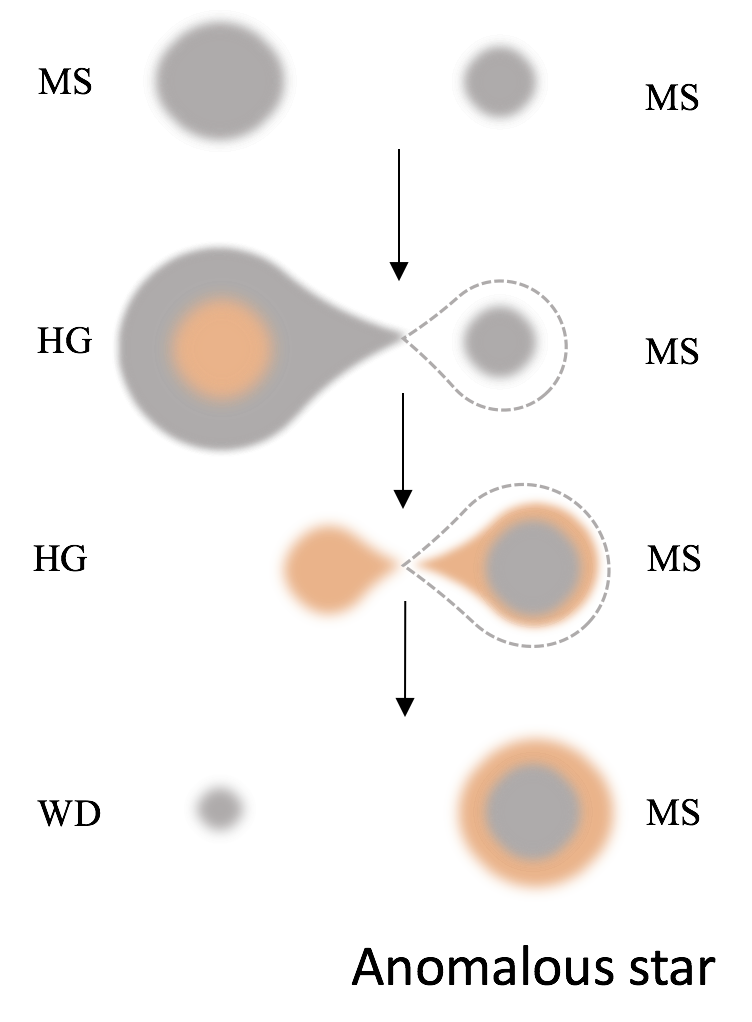}
 \caption{Schematic illustration of the stable RLOF scenario. MS, HG, and WD represent main sequence, Hertzsprung gap and white dwarf, respectively.}\label{fig:cartoon}  
  \end{center}
\end{figure}

  Fig.\,\ref{fig:cartoon} illustrates the stable RLOF scenario producing anomalous stars. The donor star (the initial more massive one) fills its Roche lobe, and starts to transfer mass to its companion star at the Hertzsprung Gap phase. The center of the donor star has already undergone H-burning process. The outer layer material of the donor star is transferred to the companion star.  If the material processed by H-burning pollutes the surface of the already formed companion star, the companion star may show abundance anomalies, which are different from the initial abundance patterns. Eventually, the donor star becomes a faint white dwarf, and the companion star may be observed as an unevolved star with abundance anomalies.

In this paper, we used the Modules for Experiments in Stellar Astrophysics code \citep[MESA, version 10398]{Paxton2011ApJS, Paxton2013ApJS, Paxton2015ApJS,2018ApJSPaxton} to carry out binary evolution, in which both stars have been calculated simultaneously. The default opacity of ``GS98'' \citep{Grevesse1998SSRv} is adopted. The nuclear network we adopt includes 45 isotopes, coupled by reactions of the pp chains, CNO, NeNa and MgAl cycles. For the initial chemical composition, we adopted a helium mass fraction of Y = 0.244, metallicity of Z = 0.0034, $\rm [O/Fe]=+0.4$ and $\rm [Na/Fe]=-0.3$, which is consistent with the observational data of 47 Tucanae.  
In our calculation, the mass transfer rate is given by \cite{Ritter1988A&A}:
\begin{equation}
\dot{M}_{\rm RLOF} \propto {\rm exp}(\frac{R_{1}-R_{L1}}{H_{p}}),
\end{equation}
where $R_{1}$ and $R_{L1}$ are the radius of the donor  star and its Roche lobe, and $H_{P}$ is the pressure scale height. 
The angular momentum loss from magnetic braking and gravitational waves are 
\begin{equation}
\dot{J}_{\rm mb} = -3.8\times10^{-30}MR_{\odot}^{4}\left(\frac{R}{R_{\odot}}\right)^{\gamma}\left(\frac{2\pi}{P_{\rm orb}}\right)^{3} 
\end{equation} 

\begin{equation}
\dot{J}_{\rm gr} = -\frac{32}{5c^{5}}\left(\frac{2\pi G}{ P_{\rm orb}}\right)^{\frac{7}{3}}\frac{\left(M_{1}M_{2}\right)^{2}}{\left(M_{1}+M_{2}\right)^{\frac{2}{3}}}
\end{equation} 

The default value of ${\gamma} = 3 $ in MESA is adopted in our calculations. We used the standard mixing length theory of convection, and the mixing length $l = \alpha H_{P}$ with $\alpha = 2$.  We treated convection mixing of elements as diffusion process. And we also set the overshooting mixing diffusion coefficient \citep{2000A&AHerwig}:
\begin{equation}
D_{\rm ov} \propto {\rm exp}\left(-\frac{2z}{fH_{p,0}}\right)
\end{equation}
Where $H_{p,0}$ is the pressure scale height at the convective boundary, and z is the distance in the radiative layer away from that location. The $f$ describes the efficiency of the extra diffusive mixing, and we adopted $ f = 0.016 $ \citep{1992A&ASSchaller}. The mass transfer was assumed to be not conservative. The mass-transfer efficiency ($\beta$) represents the fraction of transferred material which is accreted by the accretor. Because the value of the mass-transfer efficiency is still unclear in low-mass binary systems, we just chose $\beta$ = 0.5 in most cases in this paper, and we also discussed the effects of different mass-transfer efficiencies on our results in Section \ref{assumptions}. In our models, the initial donor mass ranges from 0.9 to 1.9\,$\rm {M}_\odot$. For given donor mass ($M_{1}$), various initial periods ($ P_{\rm orb}$) and mass radios ($q = M_{2}/M_{1}$) have been adopted to make the donors start mass transfer at the Hertzsprung Gap. The cases, in which the donors start mass transfer in other phases , may be studied later. The binary evolution terminates when the accretor overflows its Roche lobe. 


\section{Results}
\label{results}
\subsection{The typical binary evolution} 
\label{cases}

\subsubsection{Case 1}
\label{first}

\begin{figure*}
\begin{center}
\includegraphics[width=0.9\textwidth]{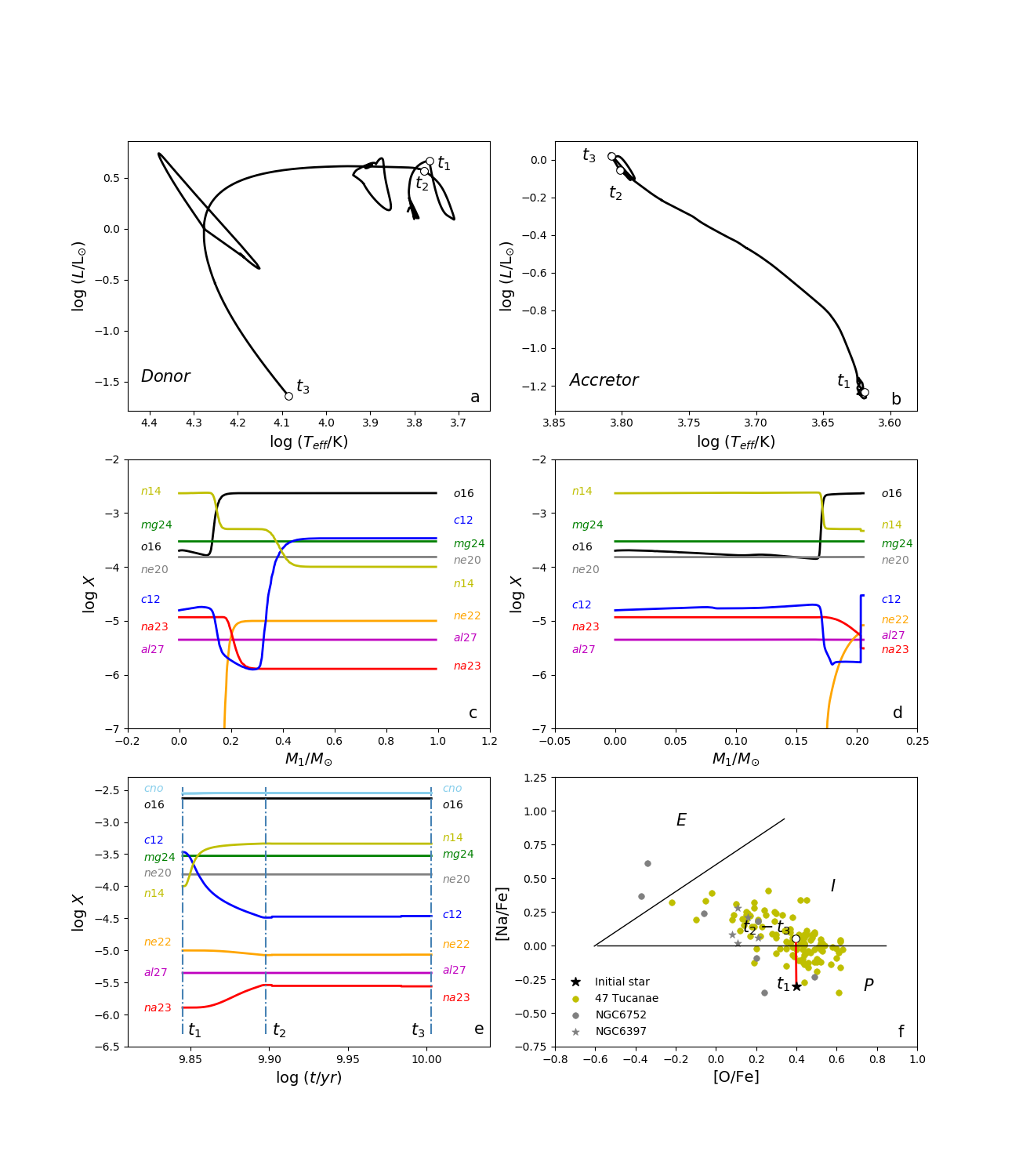} 
\caption{A binary evolution for system with $M_{1} = 1.0\,\rm {M}_\odot$, $q = 0.5$, $ P_{\rm orb} = 3.33 $\,d and $\beta = 0.5$. The panels (a) and (b) show the evolutionary tracks of both stars (the donor and the accretor). The panels (c) and (d) are the abundance profiles of the donor at the start and end of the mass-transfer phase, respectively. The panel (e) is time evolution of surface abundances of the accretor since the mass transfer occurrs. The panel (f) shows the evolution of surface Na and O abundances of the accretor, comparing with unevolved (or scarcely evolved) stars in GCs observed by \citet{Gratton2001A&A} and \citet{Dobrovolskas2014A&A}. The black lines are used to separate the primordial (P), intermediate (I) and extreme (E) populations \citep{Carretta2009aA&A}: stars with $\rm \left[O \right/Na] \leq -0.6$ dex belong to the extreme population, while those with $\rm -0.3 \leq \left[Na \right/ Fe] \leq 0.0$ are the primordial population. }\label{fig:first example}
  \end{center} 
  
\end{figure*}

\begin{figure*}
\begin{center}
\includegraphics[width=0.9\textwidth]{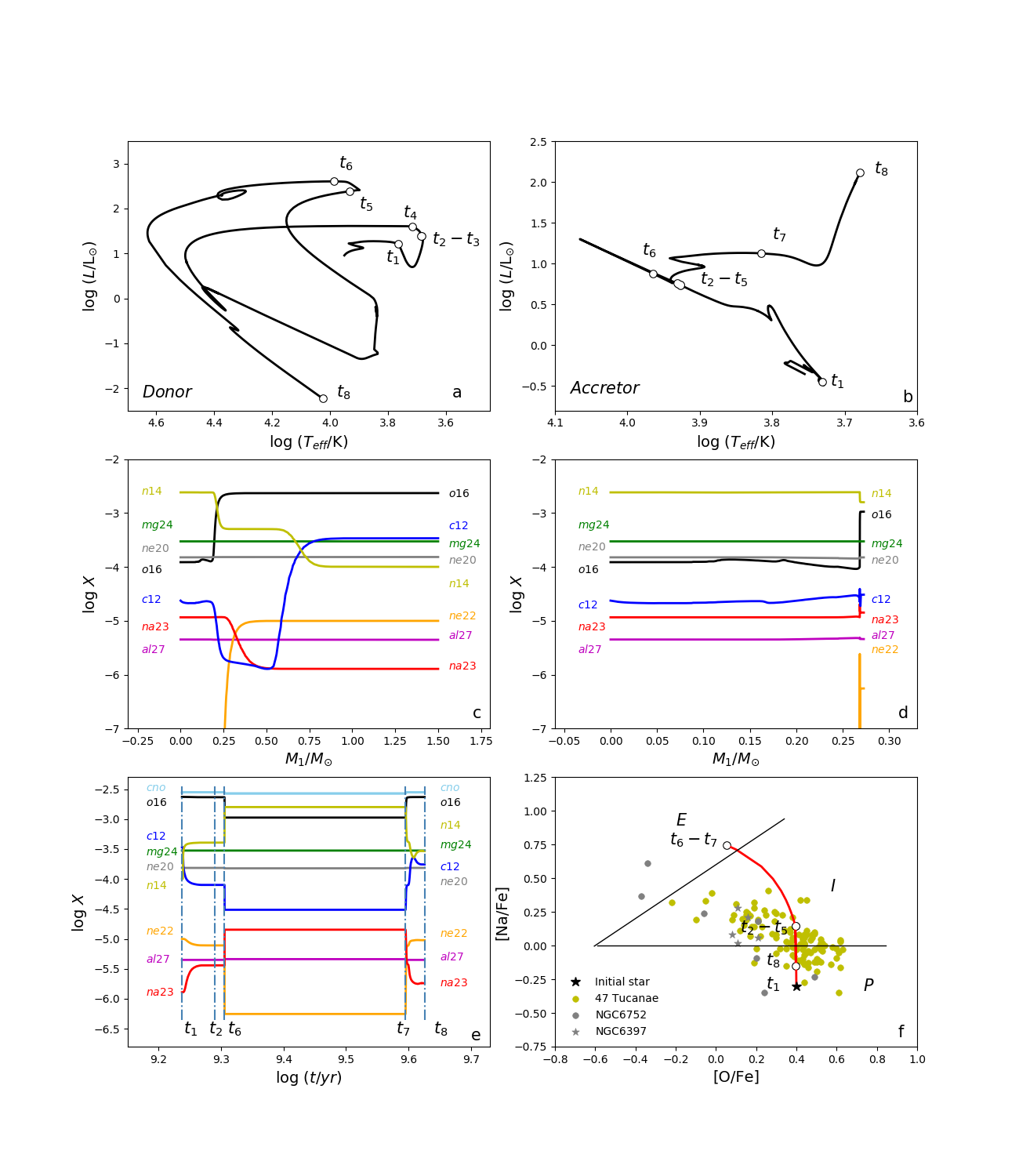} 
 \caption{Similar as Fig.\, \ref{fig:first example}, but this is a binary evolution for system with $M_{1} =1.5\,\rm {M}_\odot$, $q = 0.5$, $ P_{\rm orb} = 2.9 $\,d, and $\beta = 0.5$. The track of the accretor can be divided into several parts corresponding to various ages of the star: $t_{1}$ -- $t_{2}$, $t_{3}$ -- $t_{4}$ and $t_{5}$ -- $t_{6}$, represent the first, second and third mass-transfer phase, respectively. The envelope of the accretor becomes convective when it evolves to $t_{7}$, and $t_{8}$ means that the accretor fills its Roche lobe. The panel (f) shows the evolution of surface Na and O abundances of the accretor, from the beginning of mass transfer to the end of evolution ($t_{1} \leq t \leq t_{8}$).}\label{fig:second example} 
  \end{center}
  
\end{figure*}

\begin{figure*}
\begin{center}
\includegraphics[width=0.9\textwidth]{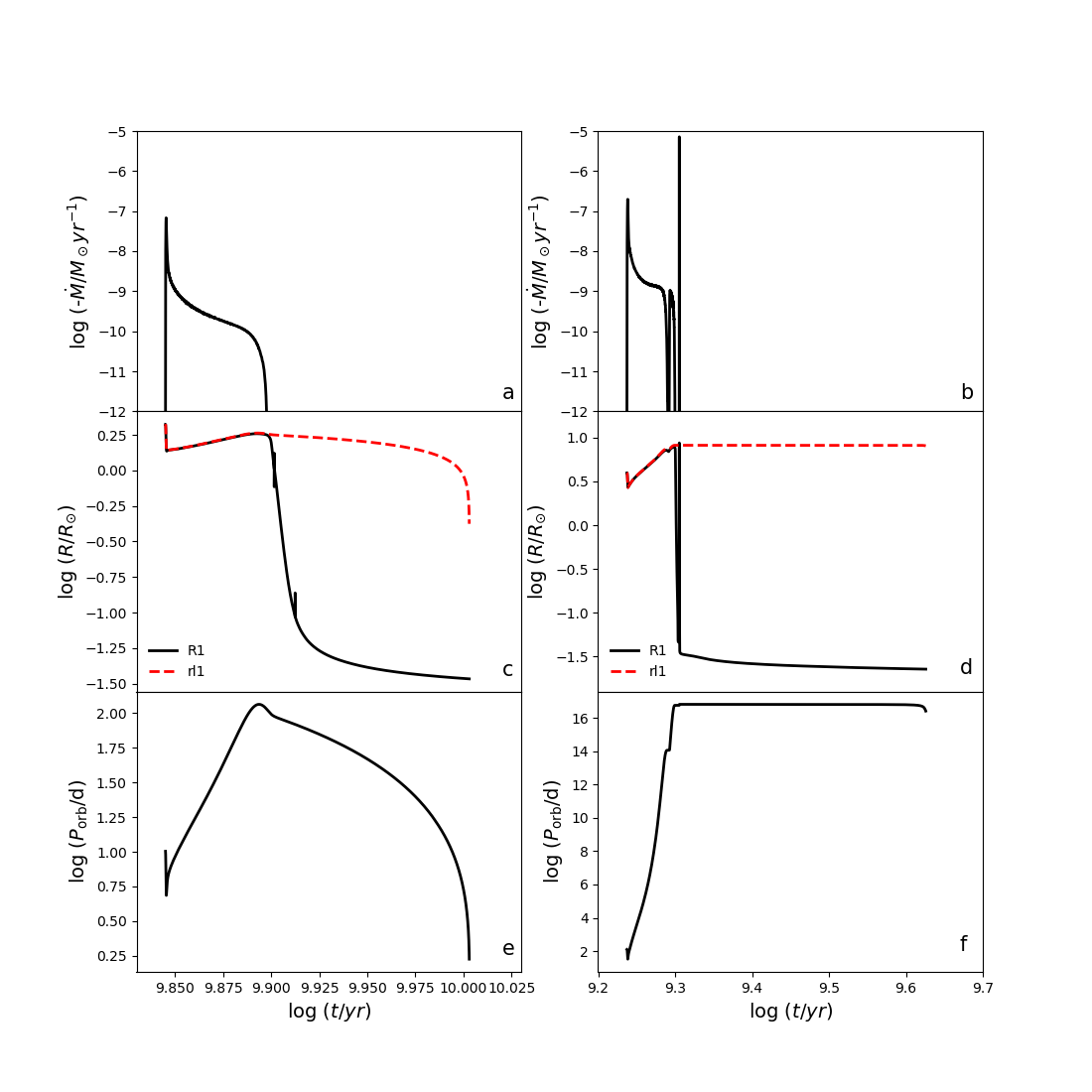} 
 \caption{The left panels (a, c, e) and the right panels (b, d, f) show the evolutionary details for the cases with 1.0\,$\rm {M}_\odot$ shown in Fig.\,\ref{fig:first example}, and 1.5\,$\rm {M}_\odot$ shown in Fig.\,\ref{fig:second example}, respectively. The panels (a) and (b) show the evolution of mass transfer rate of the donor. In the panels (c) and (d), the black solid lines (R1) show the evolution of the radii of the donors, and the red dashed lines (rl1) show the evolution of the radii of the donors' Roche lobe.  The panels (e) and (f) show the evolution of orbital periods.}\label{fig:details of examples}
  \end{center} 
  
\end{figure*}

                                                                                                                                                                                                                                                                                                                                                                                                                                                                                                                                                                                                                                                                                                                                                                                                                                                                                                                                                                                                                                                                                                                                                                                                                                                                                                                                                                                                                                                                                                                                                                                                                                                                                                                                                                                                                                                                                                                                                            \begin{table*}
	\centering
	\caption{The [O/Na] and abundances variations on the surface of the accretors relative to initial abundance, at the end of mass transfer (stable RLOF scenario). The age is the timescale of the binary systems from the beginning to the end of computation.} 
\label{tab:models}
	\begin{tabular}{l|ccccccccccccc} 
		\hline
		 Initial donor mass  & 0.9\,$\rm {M}_\odot$& 1.0\,$\rm {M}_\odot$ & 1.1\,$\rm {M}_\odot$ & 1.2\,$\rm {M}_\odot$ & 1.3\,$\rm {M}_\odot$ & 1.4 \,$\rm {M}_\odot$ & 1.5\,$\rm {M}_\odot$ & 1.6\,$\rm {M}_\odot$ & 1.7\,$\rm {M}_\odot$ & 1.8\,$\rm {M}_\odot$ & 1.9\,$\rm {M}_\odot$ \\
		\hline
		$ P_{\rm orb} ($day$)$  & 3.42 & 3.33 & 3.20 & 3.03 & 2.95 & 2.70 & 2.70 & 2.70 & 2.73 &2.80 & 3.00 \\
		\hline
		$\Delta[^{12}\textrm{C/Fe}]$ & -0.81  &-1.02 &-0.91 & -0.82 & -1.08 &-1.08 &-1.04 &-1.03 &-0.85 &-0.74 & -0.73\\
		$\Delta[^{14}\textrm{N/Fe}]$ & 0.64 & 0.66 & 0.65 &0.64 &1.19 &1.19 &1.21 &1.21 &1.32 &1.37 &1.38\\
		$\Delta[^{16}\textrm{O/Fe}]$  & $\ll$-0.01&$\ll$-0.01 &$\ll$-0.01 &$\ll$-0.01 &-0.33&-0.33 &-0.36 &-0.36 &-0.73 &-1.29 &-1.38\\
		$\Delta[^{20}\textrm{Ne/Fe}]$ & $\ll$-0.01 &$\ll$-0.01 &$\ll$-0.01 & $\ll$-0.01 & $\ll$-0.01 &$\ll$-0.01 & -0.01 &-0.01 &-0.02 &-0.02 &-0.02 \\
		$\Delta[^{22}\textrm{Ne/Fe}]$& -0.02 &-0.07 &-0.09  &-0.11 &-1.20 &-1.17 &-1.36 &-1.32&-0.94 &-1.06&-1.13\\ 
		$\Delta[^{23}\textrm{Na/Fe}]$ &  0.14& 0.35 &0.41 &0.44 & 1.00 &1.00 &1.04 &1.05 &1.12 &1.13 &1.14\\
		
		$\Delta[^{24}\textrm{Mg/Fe}]$& $\ll$0.01 &$\ll$-0.01 &$\ll$-0.01 & $\ll$-0.01 &$\ll$0.01 &$\ll$0.01&$\ll$0.01 & $\ll$0.01 & $\ll$-0.01 &-0.01 &-0.01\\
		$\Delta[^{27}\textrm{Al/Fe}]$ &$\ll$ 0.01 &$\ll$0.01 &$\ll$0.01 & $\ll$0.01 &0.01 &0.01 &0.01 &0.01 &0.05 &0.08 &0.09\\
		$\Delta$[(C+N+O)/Fe] & 0.01 & 0.01& 0.01& 0.01 &-0.02 &-0.02 &-0.02 &-0.02 &-0.03 &-0.04 &-0.04 \\
		$\rm [O/Na]$  & 0.55 & 0.35& 0.29&0.26 &-0.63 & -0.63&-0.70&-0.71 &-1.14 &-1.72&-1.82\\
		Populations  & P & I& I& I&E &E &E &E &E &E&E\\
		Age (Gyr) & 11.64 & 10.07 & 10.83 & 9.20 & 7.12 & 5.52 & 4.28 & 3.53 & 2.93 & 2.42 & 2.04 \\
		\hline
	\end{tabular}
\end{table*}

                                                                                                                                                                                                                                                                                                                                                                                                                                                                                                                                                                                                                                                                                                                                                                                                                                                                                                                                                                                                                                                                                                                                                                                                                                                                                                                                                                                                                                                                                                                                                                                                                                                                                                                                                                                                                                                                                                                                                            Fig.\,\ref{fig:first example} shows the evolution of a binary system with $M_1 = 1.0\,\rm {M}_\odot$ , $q = 0.5$, \textbf{$P_{\rm orb} = 3.33$}\,d and $\beta = 0.5$.                                                                                                                                                                                                                                                                                                                                                                                                                                                                                                                                                                                                                                                                                                                                                                                                                                                                                                                                                                                                                                                                                                                                                                                                                                                                                                                                                                                                                                                                                                                                                                                                                                                                                                                                                                                                                                                                                                                                                   Fig.\,\ref{fig:first example}(a) and (b) represent the evolutionary tracks of two stars in the Hertzsprung-Russell diagrams. The points $t_{1}$, $t_{2}$ and $t_{3}$ indicate the positions of the beginning of RLOF, the end of RLOF, and termination of the evolution, respectively. The donor star fills its Roche lobe at an age of $t_{1} = 6.99 $\,Gyr, and then transfers mass to its companion star until an age of $t_{2} = 7.90 $\,Gyr. The evolution of this binary terminates at an age of $t_{3}$, because the accretor fills its Roche lobe. At this moment, the donor star evolves into a faint white dwarf, but the companion star is still a main-sequence star but with larger mass.  

Fig.\,\ref{fig:first example}(c) and (d) indicate abundance profiles of the donor at the start and end of the mass-transfer phase, respectively. The CNO cycle and NeNa cycle have already occurred in the donor's core, when the donor begins mass transfer at Hertzsprung Gap. Mass transfer continually removes the external layers of the donor,
and the deeper chemical processed material appears at the surface of the donor. Meanwhile, the removed material from the donor is added to the outer layer of the accretor during the mass-transfer phase. Eventually, the material processed H-burning pollutes the surface of the accretor.  

Fig.\,\ref{fig:first example}(e) reveals the evolution of surface abundances of the accretor since the mass transfer begins. The dash-dot blue vertical lines represent different ages: $t_{1}$, $t_{2}$, and $t_{3}$. By the time of $t_{2} = 7.90$\,Gyr, $ ^{14}\textrm{N}$ abundance has risen by a factor of about 4.61, while the $ ^{12}\textrm{C}$ has dropped by a factor of about 10.49. The $ ^{23}\textrm{Na}$ abundance begins to rise later, and by the time of $t_{2} = 7.90 $\,Gyr, it has risen by a factor of about 2.25, while the $^{22}\textrm{Ne}$ abundance has dropped by a factor of about 1.18. The other elements ($ ^{16}\textrm{O}$, $^{20}\textrm{Ne}$, $^{24}\textrm{Mg}$, $^{27}\textrm{Al}$, and $\rm C+N+O$) show scarce variations during the mass-transfer phase. The surface abundances of the accretor roughly remain constant since the end of mass transfer. So the accretor shows C depletion, N and Na enhancement for 2.17\,Gyr ($t_{2} \leq t \leq t_{3}$), when it remains a main-sequence star.

Fig.\,\ref{fig:first example}(f) shows the comparison of the surface abundances of the accretor with the observed Na-O anti-correlation. The black star and red line represent the initial location and the evolution of surface abundances of the accretor, respectively. At the end of mass-transfer phase ($t= t_{2}$), the accretor shows a maximum surface $\rm Na^{23}$. The accretor is consistent with the I population with a timescale of 2.17\,Gyr  ($t_{2} \leq t \leq t_{3}$), when it remains a main-sequence star.

\subsubsection{Case 2}
 \label{second}
     We also give another example of binary evolution with $M_1 = 1.5\,\rm{M}_\odot$, $q = 0.5$, $P_{\rm orb} = 2.9$\,d and $\beta = 0.5$ in Fig.\,\ref{fig:second example}. The main difference between this example and the previous one is that this binary system has undergone three mass-transfer episodes, as indicated by $t_{1}$ -- $t_{2}$ , $t_{3}$ -- $t_{4}$ and $t_{5}$ -- $t_{6}$, respectively. The second mass transfer has little effect on the surface abundances of the accretor. Finally, the donor evolves into a faint white dwarf, and the accretor becomes a RGB star.

At the end of the third mass-transfer episode (t = $t_{6}$) as shown in Fig.\,\ref{fig:second example}(e), the surface abundances of $^{14}\textrm{N}$ and $^{23}\textrm{Na}$ of the accretor increase by a factor of 15.80 and 11.07, respectively, while that of  $^{12}\textrm{C}$, $^{16}\textrm{O}$ and $^{22}\textrm{Ne}$ decrease by a factor of 11.13, 2.21 and 17.94, respectively. The other elements abundances ($^{20}\textrm{Ne}$, $^{24}\textrm{Mg}$, $^{27}\textrm{Al}$, and $\rm C+N+O$ ) remain nearly unchanged. 

After the end of the third mass-transfer episode, the surface abundances of the accretor remain constant until the convection of envelope occurs at an age of $t_{7}$ = 3.93\,Gyr. The convection from the outside to the inside changes the abundances of envelope in a short timescale (about 0.29\,Gyr, from  $t_{7}$ to $t_{8}$). During this process, the surface abundances of $^{23}\textrm{Na}$ and $^{14}\textrm{N}$ reduce rapidly, while that of $^{12}\textrm{C}$ and $ ^{22}\textrm{Ne}$ rise quickly. Eventually, the surface $^{14}\textrm{N}$ and $^{23}\textrm{Na}$ rise by a factor of about 2.92 and 1.41, respectively, and the $^{12}\textrm{C}$ drops by a factor of about 1.95, while the other elements ($^{16}\textrm{O}$,$^{20}\textrm{Ne}$, $^{22}\textrm{Ne}$,$^{24}\textrm{Mg}$, $^{27}\textrm{Al}$,  and $\rm C+N+O$) show scarce variations, compared to the initial abundances. So the accretor shows C-N anti-correlation and Na-O anti-correlation for 1.91\,Gyr ($t_{6} \leq t \leq t_{7}$), when it remains a main-sequence star, main-sequence turn-off star or early subgiant star. However it shows a weaker C-N anti-correlation, and scarcely Na enhancement after the convection of envelope occurs. 
 
Fig.\,\ref{fig:second example}(f) shows that the surface abundances of the accretor contribute to E population at the end of the third RLOF, lasting for a long timescale of 1.91\,Gyr ($t_{6} \leq t \leq t_{7}$). But the surface abundances will be consistent with the P population after the convection of envelope takes place. Therefore, the accretor is an anomalous star belonging to E population when it evolves into a main sequence star, main-sequence turn-off star or early subgiant star. As long as the accretor becomes a red giant branch star, it does not reveal abundances anomalies. 

\subsubsection{The binary evolution details of these two cases}
\label{evolutionary details}
We also explored the differences of the two aforementioned cases in mass-transfer rate, orbital period, radius of the donors and radius of donors' Roche lobe, since the mass transfer begins, as shown in Fig.\,\ref{fig:details of examples}. There is only one episode of the RLOF in the first case with initial parameters of $M_1 = 1.0\,\rm {M}_\odot$, $q = 0.5$ and $ P_{\rm orb}= 0.33\rm\,d$. The mass transfer begins, when the donor fills its Roche lobe, and the mass-transfer rate rises to $10^{-7}$\,$\rm {M}_\odot \rm yr^{-1}$ quickly. The mass transfer ceases as the donor contracts and deviates from its Roche lobe. The orbital period decreases at the beginning of the mass-transfer phase, and then increases when the mass of the accretor is larger than that of the donor. The orbital period decreases after the mass-transfer phase because of the large loss of angular momentum. 

There are three episodes of the RLOF in the second case with initial parameters of $M_1 = 1.5\,\rm {M}_\odot$, $q = 0.75$ and $ P_{\rm orb}= 2.90$\,d. When the H-burning shell arrives at the hydrogen discontinuity, the donor contracts, and then the first episode comes to end. The second episode occurs, when the the H-burning shell passes through the hydrogen discontinuity. During the second episode, the mass-transfer rate is not large, the maximum of which is $\sim$ $10^{-9}\,\rm {M}_\odot \rm yr^{-1}$. The third episode begins when the donor significantly expands due to a further burst of hydrogen burning (H-flash). The mass-transfer rate of the third episode can even reach $10^{-5.2}\,\rm {M}_\odot \rm yr^{-1}$. The third mass-transfer episode has a significant effect on the surface abundances of the accretors.

\begin{figure}
\begin{center}
\includegraphics[width=0.5\textwidth]{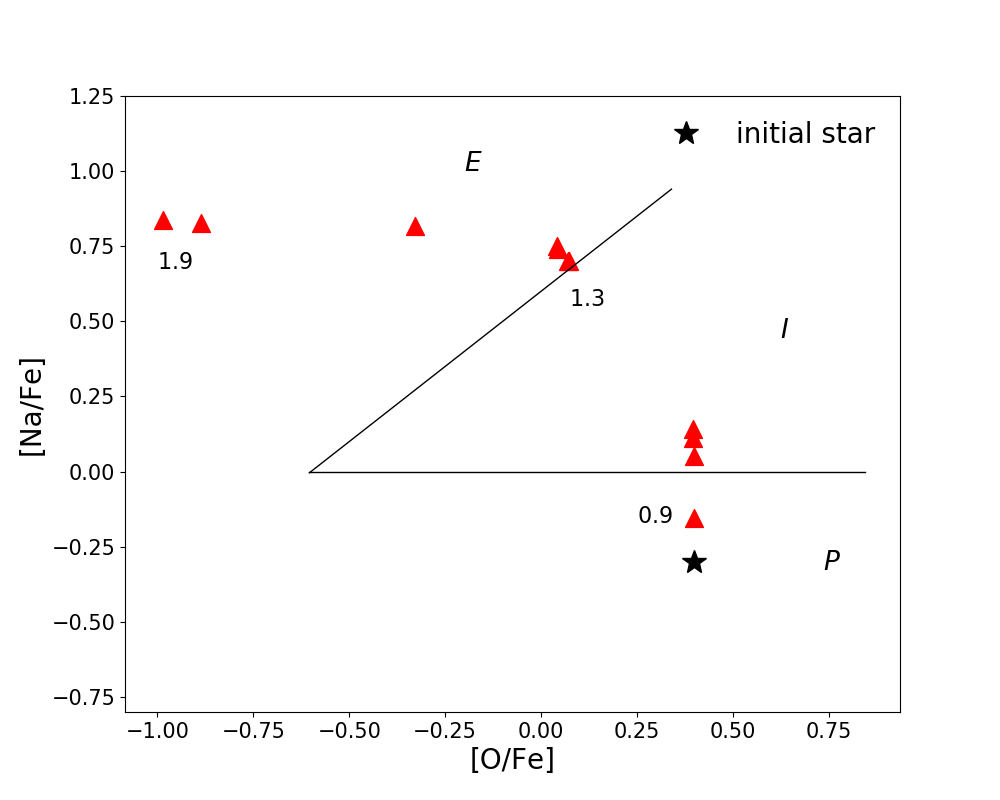} 
 \caption{The comparison of binary systems with different donor masses. In these models, the mass of donor star is between $0.9\,\rm {M}_\odot$ and $1.9\,\rm {M}_\odot$, $q= 0.5$  and $\beta = 0.5$. The triangles indicate the surface abundances of the accretors at the end of mass transfer, when they are still main-sequence stars. We use the black lines to separate the primordial (P), intermediate (I) and extreme (E) populations. }\label{fig:mass} 
  \end{center}
  
\end{figure}

\begin{figure*}
\begin{center}
\includegraphics[width=1\textwidth]{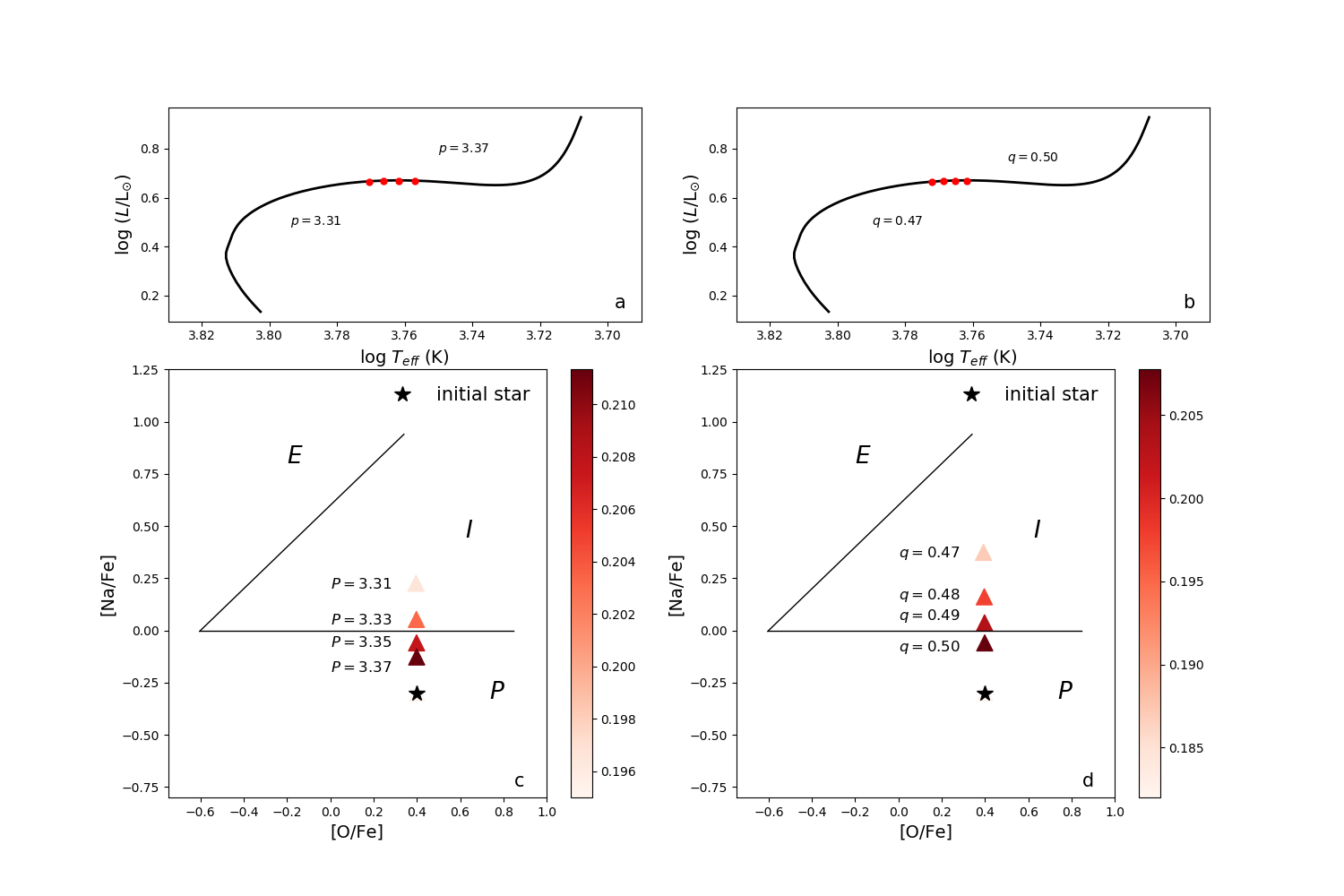}
 \caption{The comparison of binary systems with different parameters (mass ratios and initial orbital periods). The triangles indicate the surface abundances of the accretors at the end of mass transfer, when they are still main-sequence stars. In the panel (c), the binary systems have $M_{1} = 1.0\,\rm {M}_\odot$, $q = 0.5$, and $\beta = 0.5$ but with different initial orbital periods (from the top to the bottom: 3.31, 3.33, 3.35, 3.37\,d). The dots on the Hertzsprung-Russell diagram in panel (a) indicate the start of mass transfer (from the left to the right: 3.31, 3.33, 3.35, 3.37\,d). In the panel (d), the binary systems have the donor mass of $M_{1} = 1.0 \,\rm {M}_\odot$, initial orbital period of $P_{\rm orb} = 3.38$\,d, but with different mass ratios (from the top to the bottom: 0.47, 0.48, 0.49, 0.50). The red dots on the Hertzsprung-Russell diagram in panel (b) indicate the start of mass transfer for binaries with different mass ratios (from the left to the right: 0.47, 0.48, 0.49, 0.50). The color bar represents the mass of donor at the end of mass transfer.}\label{fig:pq}
  \end{center}
  
\end{figure*}

\begin{figure}
\begin{center}
\includegraphics[width=0.5\textwidth]{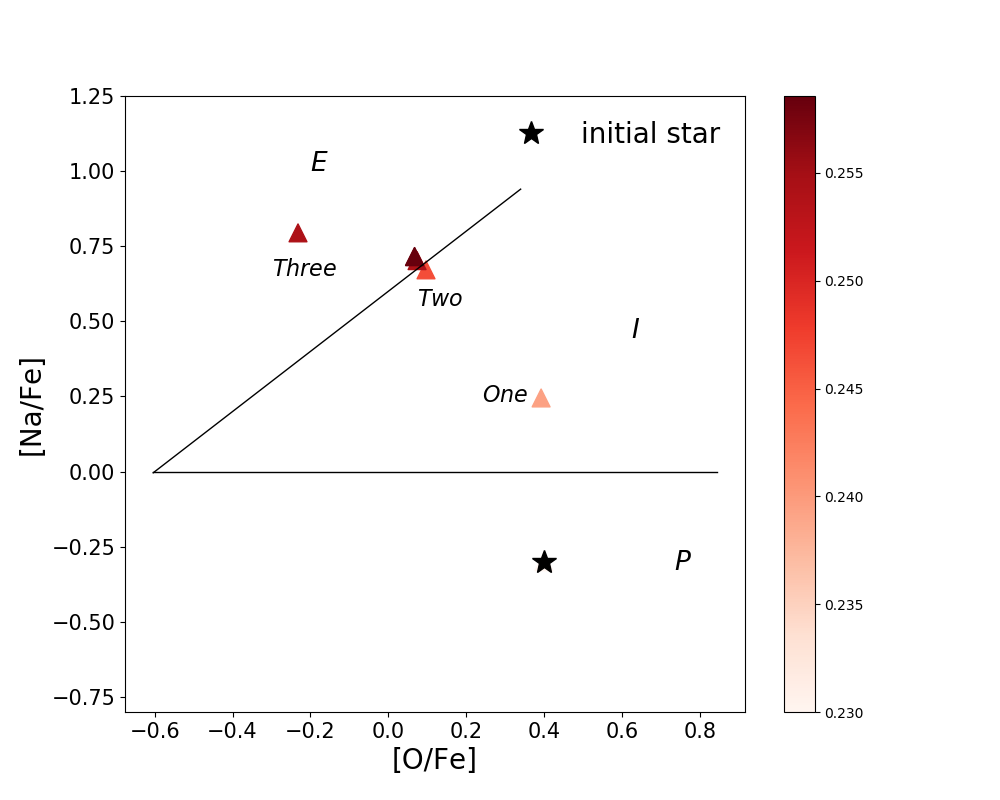}
 \caption{Comparison of binary systems with donor mass of $M_{1} = 1.4\, \rm {M}_\odot$, but different initial orbital periods ($ P_{\rm orb}$ = 2.55, 2.6, 2.7, 2.73, 2.78, 2.8\,d). The binary system with $ P_{\rm orb}$ = 2.55\,d only have one mass-transfer episode. The binary system with $ P_{\rm orb}$ = 2.73\,d have three mass-transfer episodes ($\dot{M}$ $\geq$ $10^{-8}\,\rm {M}_\odot \rm yr^{-1}$), and other systems have two mass-transfer episodes. The triangles indicate the surface abundances of the accretors at the end of mass transfer, when they are still main-sequence stars. The color bar represents the mass of donor at the end of mass transfer.}\label{fig:another P}     
  \end{center}
  
\end{figure}

\begin{figure}
\begin{center}
\includegraphics[width=0.5\textwidth]{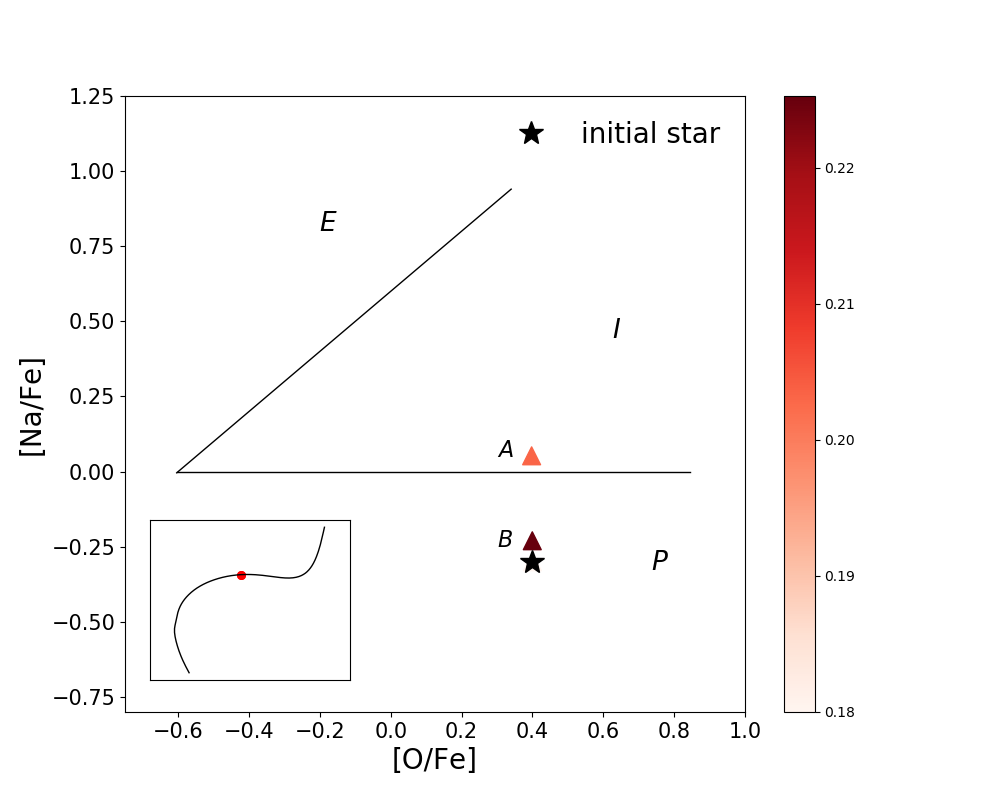} 
 \caption{The effect of magnetic braking on the surface abundances of the accretors. Both binary systems have  $M_{1} = 1.0\, \rm {M}_\odot$, $q = 0.5$ and $\beta = 0.5$. The binary system A ($ P_{\rm orb}$ = 3.330\,d) and B ($ P_{\rm orb}$ = 1.026\,d) are calculated with and without magnetic braking, respectively. The triangles indicate the surface abundances of the accretors at the end of mass transfer. The color bar represents the mass of donor at the end of mass transfer. Both binaries begin mass transfer at same location in the HR diagram (red dot).}\label{fig:magnetic}
  \end{center}

\end{figure}

\begin{figure*}
\begin{center}
\includegraphics[width=1\textwidth]{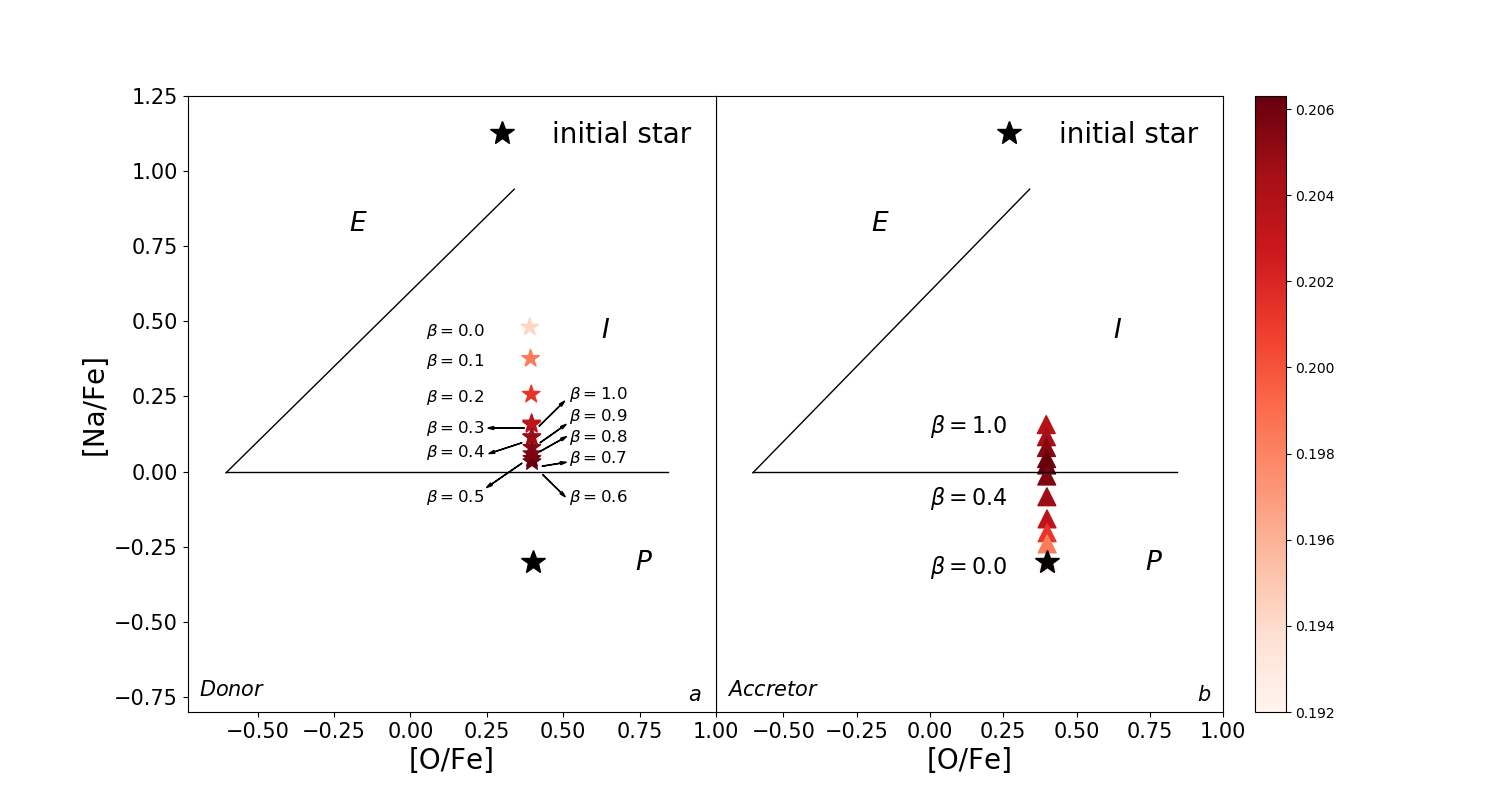} 
 \caption{The effect of mass-transfer efficiency ($\beta$) on the surface abundances of donor stars in panel (a), and the accretors in panel (b). The binary systems with the donor mass  $M_{1} = 1.0\,\rm {M}_\odot$, mass ratio $q = 0.5$, and $ P_{\rm orb}= 3.34$\,d but with different $\beta$ (from bottom to the top in panel (b): 0.0 -- 1.0, in steps of 0.1; $\beta$ in panel (a) is already marked). The red stars and triangles indicate the surface abundances of the donor stars and the accretors at the end of mass transfer, respectively. The color bar represents the mass of donor at the end of mass transfer in both panel (a) and (b).}\label{fig:beta}
  \end{center}
  
\end{figure*}

\subsection{The influence of model parameters and assumptions}
\label{influence}

 Here we are interested in the effects of the parameters ($M_{1}$, $q = M_{2}/M_{1}$, $ P_{\rm orb}$), and some assumptions, such as  $\beta$ and magnetic braking, on the surface abundances of the accretors. So we calculated the binary systems with different parameters and assumptions. In these models, We focus on the surface abundances of the accretors at the end of mass transfer, when they are still non-evolved or scarcely evolved stars. 

\subsubsection{The influence of model parameters}
\label{parameters}
Fig.\,\ref{fig:mass} shows surface abundances of the accretors in binaries with different initial donor masses. The binary systems with larger donor mass reproduce anomalous accretors with higher $^{23}\textrm{Na}$ enhancement, even the $^{16}\textrm{O}$ poor. The surface abundances of the accretors in binary systems with $M_{1} \geq 1.3\,\rm {M}_\odot$ are consistent with the extreme population due to multiple mass-transfer episodes. The binary systems with $1.0\,\rm {M}_\odot \leq M_{1} \leq 1.2\,\rm {M}_\odot$ contribute to intermediate population. The details of abundances are showed in Table \,\ref{tab:models}. In these binary systems, N and Na increase, while C and even O decrease. It is obvious that  $^{24}\textrm{Mg}$, $^{27}\textrm{Al}$, and $\rm C+N+O$ remain constant. The binary systems with larger donor mass ($M_{1} \geq 1.2\,\rm {M}_\odot$ ) have shorter timescale, and they show more significant abundance anomalies. The binary systems with $M_{1} \leq 1.1\,\rm {M}_\odot$ partly contribute to the abundance anomalies of old GCs ($\geq 10$\,Gyr).

Fig.\,\ref{fig:pq} shows the effects of different initial orbital periods and mass ratios on the surface abundances of the accretors at the end of mass transfer. In the binary systems with smaller mass ratio, or shorter initial orbital period, the donor begins mass-transfer earlier, and more material is transferred to the companion star. In the end, the accretor shows more pronounced $ ^{23}\textrm{Na}$ enhancement. 

All of these consequences are completed under the condition that the masses of the donors in these binary systems are 1.0\,$ \rm {M}_\odot$. It will be different if the binary system undergoes more than one mass-transfer episode. Fig.\,\ref{fig:another P} shows the effects of different initial orbital periods in binary systems with $M_{1} = 1.4\,\rm {M}_\odot$ and $q = 0.7$. They do not show higher $^{23}\textrm{Na}$ abundance, even though the binary systems with shorter initial orbital period transfer more material to the accretors. On the contrary, the binary systems with larger initial orbital period exhibit higher $^{23}\textrm{Na}$ abundance and even lower $^{16}\textrm{O}$ abundance, due to  multiple mass-transfer episodes. So our results indicate that the number of mass-transfer episodes influences the degree of abundance anomalies.

\subsubsection{The influence of model assumptions}
\label{assumptions}

 Fig.\,\ref{fig:magnetic} shows the effect of magnetic braking on the surface abundances of the accretors. Both binary systems have the same donor mass and the same mass ratio. Besides, the donors of the two systems begin mass transfer at the same position on their Hertzsprung-Russell diagrams. The magnetic braking takes away lots of angular momentum of the binary system, leading to a smaller orbital period. Therefore, the donor losses more material in the binary system with magnetic braking (binary system A), and the surface of the accretor shows higher $^{23}\textrm{Na}$ enhancement.

 Fig.\,\ref{fig:beta}(a) reveals the surface abundances of the donor stars at the end of mass transfer for binaries with various $\beta$. The residual donor mass after mass transfer is different with various $\beta$. This is because different $\beta$ results in different evolution of orbital period and mass ratio since mass transfer begins, which affects the amount of mass transferred. It is clear that the smaller the residual mass of the donor star is, the higher the sodium abundance of the donor star is, due to more material transferred.
  
  Fig.\,\ref{fig:beta}(b) shows the surface abundances of the accretors at the end of mass transfer for binaries with various $\beta$. The accretors have higher surface $^{23}\textrm{Na}$ abundance in binary systems with larger $\beta$. The surface abundances of the accretors depend on the abundances of material transferred from the donor stars, and the dilution on the surface of the accretors. The surface abundances of the donors at the end of mass transfer are shown in the Fig.\,\ref{fig:beta}(a), which are not exactly the same as that of the accretors even with the same $\beta$ due to dilution on the surface of the accretors. The accretor does not accrete any material when $\beta = 0.0$, so it keeps the same surface abundances with the initial star. Despite the high $^{23}\textrm{Na}$ abundance (consisting with the intermediate population) of the transferred material in the binary systems with $\beta = 0.1 \sim 0.4$, the accretors still do not show significant abundance anomalies (consistenting with the primordial population). However, the surface abundances of the accretors are almost the same as that of the donors in the binaries with $\beta = 0.5 \sim 1.0$. The reason of different dilution effects above in binary systems with various $\beta$ is that the envelop convection will gradually fade away, as the mass of the main-sequence star (the accretor) increases. With the increase of material accreted, there is no obvious dilution effect on the surfaces of the accretors in the binaries with $\beta = 0.5 \sim 1.0$ due to the fading of the envelop convection. Therefore, the two cases in section \ref{cases} do not show significant dilution with $\beta = 0.5$.

\section{Discussion and conclusions}
\label{discussion}

As an important step in exploring the contribution of the stable RLOF scenario to abundance anomalies in GCs, we studied the evolution of low-mass binaries to reproduce anomalous stars. Our models show that the stable RLOF scenario in low-mass binary systems is able to reproduce main-sequence stars or early subgiant stars with peculiar chemical patterns, e.g. C and O depletion, Na and N enhancement, constant Mg, Al and C+N+O. By comparing with observed Na-O anti-correlation in GCs, we found that the abundance patterns of the accretors can match the observed populations with different abundances (e.g. primordial, intermediate, and extreme population ).

The surface abundances of the accretors are directly related to the abundance profiles of the donor stars, the amount of material transferred, and the dilution on the accretors. Our results show that the initial parameters of the binary system (${M}_{1}$, $q = {M}_{2}/{M}_{1}$ and ${P}_{\rm orb}$), and various assumptions (mass-transfer efficiency and magnetic braking) play important roles in the abundance anomalies of the accretors. The surface of the accretors shows a stronger enhancement of Na in binary systems with more massive donors, lower mass ratio and shorter initial orbital period under the same mass-transfer episodes. This is because the nuclear reactions in the center of the massive donors are more intense due to the higher temperature compared with the lower - mass stars, which affects the abundance profiles of the donor stars. Besides, according to \citet{2000MNRASHan}, the donor stars transfer more material if the binary systems have lower mass ratio and shorter initial orbital period, which will cause higher surface $^{23}\textrm{Na}$ abundance.

The accretors in binary systems with magnetic braking and larger mass-transfer efficiency show stronger $^{23}\textrm{Na}$ enhancement. It is because the magnetic braking of the lost material carry away the orbital angular momentum, which leads to that more material are transferred. The material lost from the binary system will affect the separation and mass ratio of the systems, then the amount of material transferred vary with different mass-transfer efficiencies. Besides, the dilution effect on the surface of the accretors depends on the initial accretor mass and the mass-transfer efficiency. Only in binary systems with mass-transfer efficiency between 0.0 and 0.4 in our cases (${M}_{2} =  0.5\,\rm {M}_\odot$), the dilution effect on the accretor can significantly reduce sodium abundance. Besides, our results indicate the extra mass-transfer episodes can produce strong variations of surface abundances of the accretors, even with a little material accreted (see Fig.\,\ref{fig:another P}), because the abundance profiles of the donor stars are different during each mass-transfer episode.

Our results reveal that the stable RLOF scenario can establish the abundance variations of C, N, O and Na, but not  Mg and Al. Therefore, this scenario is at least partly responsible for some Na-rich, O-poor stars without the changes in Mg and Al. Another possibility is that the reaction rate of MgAl circle should be increased to reproduce this anti-correlation \citep{Decressin2007A&A}. It should be noted that the Mg-Al anti-correlation is not observed in all GCs, but the Na-O and C-N anti-correlations seem to be found in all GCs \citep{Bastian2018ARA&A}. In the stable RLOF scenario, the binary systems with more massive donor ($M_{1} \geq 1.2\,\rm {M}_\odot$) can contribute to the abundance anomalies in the young clusters, while those with less massive donor can contribute in the old clusters (see Table \,\ref{tab:models}).  Different chemical patterns may be formed through different scenarios, and various combinations of different scenarios (including the stable RLOF scenario) are required to explain the formation of multiple stellar populations in different GCs.

Compared with the self-enrichment scenario, the material processed H-buring in the stable RLOF binary scenario pollutes the surface of the already formed companion star through the mass transfer, and it does not require the hypothesis of extra multiple epochs of star formation. In addition, most previous self-enrichment scenarios argued the contributions of stars with masses in the range of $2.5\,\rm {M}_\odot$ -- ${8\,\rm {M}_\odot}$, or even more massive stars as candidates of polluters \citep{Decressin2007A&A, Decressin2009A&A, Mink2009A&A, 2010MNRASD'Ercole, 2012A&AVanbeveren,
Denissenkov2014MNRAS}. However, our results show that the low-mass stars with  $M<$ $2.0\,\rm {M}_\odot$ also give a possible contribution to the abundance anomalies, and probably alleviates the mass budget problem of multiple generation scenarios to a certain extent.

In the stable RLOF scenario, the anomalous star would be more massive than the normal, single star in the same evolutionary stages (e.g. the same position on the CMD). Considering the presence of binaries in field stars and open cluster, similar abundance anomalies should also exist in field stars and open cluster. Actually, some metal-poor field stars were found to be enriched in nitrogen \citep{Tang2018arXiv}, and these field stars, includes CH star, may have binary origin \citep{McClure1984ApJ,McClure1990ApJ}. A more detailed comparison of abundance patterns in different environments (e.g. the field stars, open clusters and GCs) may help to understand the binary mass transfer scenario, and their possible contribution to the abundance anomalies in GCs well.

\section*{Acknowledgements}
We are grateful to the anonymous referee, for his or her valuable suggestions which helped to improve the paper greatly. We thank Zhanwen Han, Yan Li, Heran Xiong, Xiaobo Gong, Jiao Li and Chengyuan Wu for their helpful discussions.  We are grateful to the Natural Science Foundation of China (Nos 11573061, 11661161016, 11673059, 11873085, 11733008, 11703081, 11521303, 11773065), the Natural Science Foundation of Yunnan Province (2015FB190), and Youth Innovation Promotion Association of Chinese Academy of Sciences(grant no. 2018076) for support.










\bsp	
\label{lastpage}
\end{document}